\documentclass[preprint,amsmath,aps,prb]{revtex4}
\usepackage[pdftex]{color,graphicx}
\usepackage{graphicx}
\usepackage{amsmath}
\begin{document}
\title{Effects of noise on hysteresis and resonance width in graphene and nanotubes resonators.}
\author{O.G. Cantu~Ros$^{1,2,3}$, G.~Platero$^3$, L.L.Bonilla$^{1,2}$}
\affiliation{$^1$G. Mill\'an Institute, Fluid Dynamics, Nanoscience and Industrial
Mathematics, Universidad Carlos III de Madrid, Avda.\ Universidad 30; E-28911 Legan\'es, Spain\\
$^2$Unidad Asociada al Instituto de Ciencia de Materiales de Madrid, CSIC, 28049 Cantoblanco, Madrid, Spain\\
$^3$Instituto de Ciencia de Materiales de Madrid, CSIC, 28049 Cantoblanco, Madrid, Spain}
\date{\today}
\begin{abstract}
We investigate the role that noise plays in the hysteretic dynamics of a suspended nanotube or a graphene sheet subject to an oscillating force. We find that not only the size but also the position of the hysteresis region in these systems can be controlled by noise. We also find that nano-resonators act as noise rectifiers: by increasing the noise in the setup, the resonance width of the characteristic peak in these systems is reduced and, as a result, the quality factor is increased.
\end{abstract}
\maketitle
\section{Introduction}\label{sec:1}
Recent theoretical \cite{lifshitz,lasagne} and experimental work \cite{Alm1,Badzey,Alm2,BoA,Bunch,Bach} have shown that reducing the dimension and the size of a material with a vibrational degree of freedom enhances the role of the nonlinearities in the dynamics. Theory and experiments have demonstrated noise squeezing\cite{Alm1} and stochastic resonance\cite{Badzey,Alm2,Dykman} in nanomechanical resonators. These systems are used as precision measurement devices for applications such as mass and force sensing,\cite{Aldridge} quantum motion detection and radiofrequency signal processing.\cite{Eriksson} Resonators consisting of suspended nanotubes or graphene sheets can be used as extremely accurate mass sensors with yoctogram resolution.\cite{Chaste} Models based on driven Duffing oscillators with nonlinear dissipation are used to interpret measurements in these devices.\cite{Bunch,Bach} Nonlinear terms in the model are responsible for bistability in the oscillatory response
 as a function of the driving frequency and of quality factors dependent on the driving force.

Motivated by experiments on nanotube and graphene resonators,\cite{Bunch,Bach,Chaste} we study in this work the dynamics of graphene and nano tube resonators based on a suspended wired subject to an external oscillating driving force. We take into account the nonlinear damping term, the Duffing nonlinearity and we add white noise forcing not considered in the model used to interpret the experimental results. Given the use of these resonators resonators as accurate mass sensing devices, it is important to include in the dynamics all major effects in order to describe more accurately both, the system characteristics and the different quantities inferred from it, such as, mass, quality factor, dissipation coefficients, etc.

Noise is always present in a physical system. For linear systems, noise is commonly regarded as having a destructive but relatively innocuous effect, blurring our view of a system but having no effect on the underlying processes involved. In nonlinear systems, a driving white noise term can modify drastically the deterministic dynamics. It can shift bifurcation points or induce behaviors with no deterministic counterpart, thereby affecting the dynamics and the quantities measured or inferred indirectly from the experimental measurements. Previous theoretical work showed that the main sources of noise in linear simple harmonic oscillators are thermomechanical noise, temperature fluctuations and adsorption-desorption noise.\cite{Cleland} The analysis of the sources of noise in nonlinear resonators is beyond the scope of this work. However we will consider two different cases. On the one hand, we shall assume that the sources are similar to those of linear resonators and discuss
  the general case of additive white noise, without worrying about the causes of it. On the other hand, a local nonlinear damping (dissipation) means that we may need adding an appropriate white noise force to the equations according to the fluctuation-dissipation theorem. This will result in a special type of multiplicative noise. In both cases, considering the effect of noise will allow to estimate the values of the cubic nonlinearity with higher precision than in the noiseless case. Our results may also shed some light for other possible applications.

The rest of the paper is as follows. In Section \ref{sec:2} we give a general discussion of the characteristics of nonlinear resonators, focusing on the role that the nonlinear dissipation coefficient plays on the resonator dynamics. In Section \ref{sec:3}, we add an external white noise of fixed strength to the resonator system of Section \ref{sec:2} and discuss the subsequent modifications of its main characteristics. In Section \ref{sec:4}, we discuss the effect produced by an internal noise satisfying the fluctuation-dissipation theorem. Finally, in Section \ref{sec:5} we present the main results of this work and the implications that they have in experimental measurements.

\section{Theory}\label{sec:2}
We first discuss the properties of mechanical resonators based on a suspended nano\-wire or graphene strip that is doubly clamped and subjected to an oscillating external force with frequency $f$. Since we only care about the dynamics of the fundamental mode, we can consider the graphene membrane as a one dimensional object. In this regime the only difference between the nanowire and the graphene strip is given by the parameter values of each one.\cite{Bach}

As has already been shown experimentally,\cite{Bach} the dynamics of these resonators is highly nonlinear, as they present a force-dependent quality factor and, in some cases, hysteresis in the oscillation amplitude as a function of the driving frequency. The resonator dynamics can be described by the following equation of motion\cite{Nayfeh}
\begin{equation}
m\ddot {\tilde{x}}=-k\tilde{x}-\alpha \tilde{x}^3-\gamma\dot{\tilde{x}}-\tilde{\eta} \tilde{x}^2 \dot{\tilde{x}} + F_{drive}\cos(2\pi ft),\label{eq1}
\end{equation}
where $m$ is the effective mass, $\alpha$ is the Duffing term, $\gamma\dot{\tilde{x}}$ and $\tilde{\eta} \tilde{x}^2 \dot{\tilde{x}}$ are the linear and nonlinear damping terms, and $(k+\alpha x^2)$ is the amplitude dependent spring's stiffness with a spring constant $k=m\omega_0^2$ ($\omega_0\approx 2\pi f$ is the natural frequency of the resonator fundamental mode, which is close to resonance). Depending on the sign of $\alpha$, the Duffing term modifies the stiffness of the resonator (more stiff for $\alpha>0$, softer for $\alpha<0$). Equivalently, we can work with the nondimensional equation
\begin{equation}\label{dimless}
\begin{split}
\ddot {x}=&-x-Q^{-1}\dot {x}-x^3-\eta x^2\dot {x} + F_D\cos(\Omega t)\\
\mathrm{with}&\\
x=&\tilde{x}\sqrt{\alpha m\omega_0^2},\quad t=\omega_0\tilde{t},\quad  Q^{-1}=\frac{\gamma}{m\omega_0},\quad \eta=\frac{\tilde{\eta}\omega_0}{\alpha},\quad F_D=\frac{F_{drive}}{\omega_0^3}\sqrt{\frac{\alpha}{m^3}},\quad \Omega=\frac{{2\pi f}}{\omega_0},
\end{split}
\end{equation}
which will be used all throughout the paper. This means that all the quantities appearing in this work are dimensionless except when specified otherwise.

\begin{figure}
\includegraphics[scale=0.5]{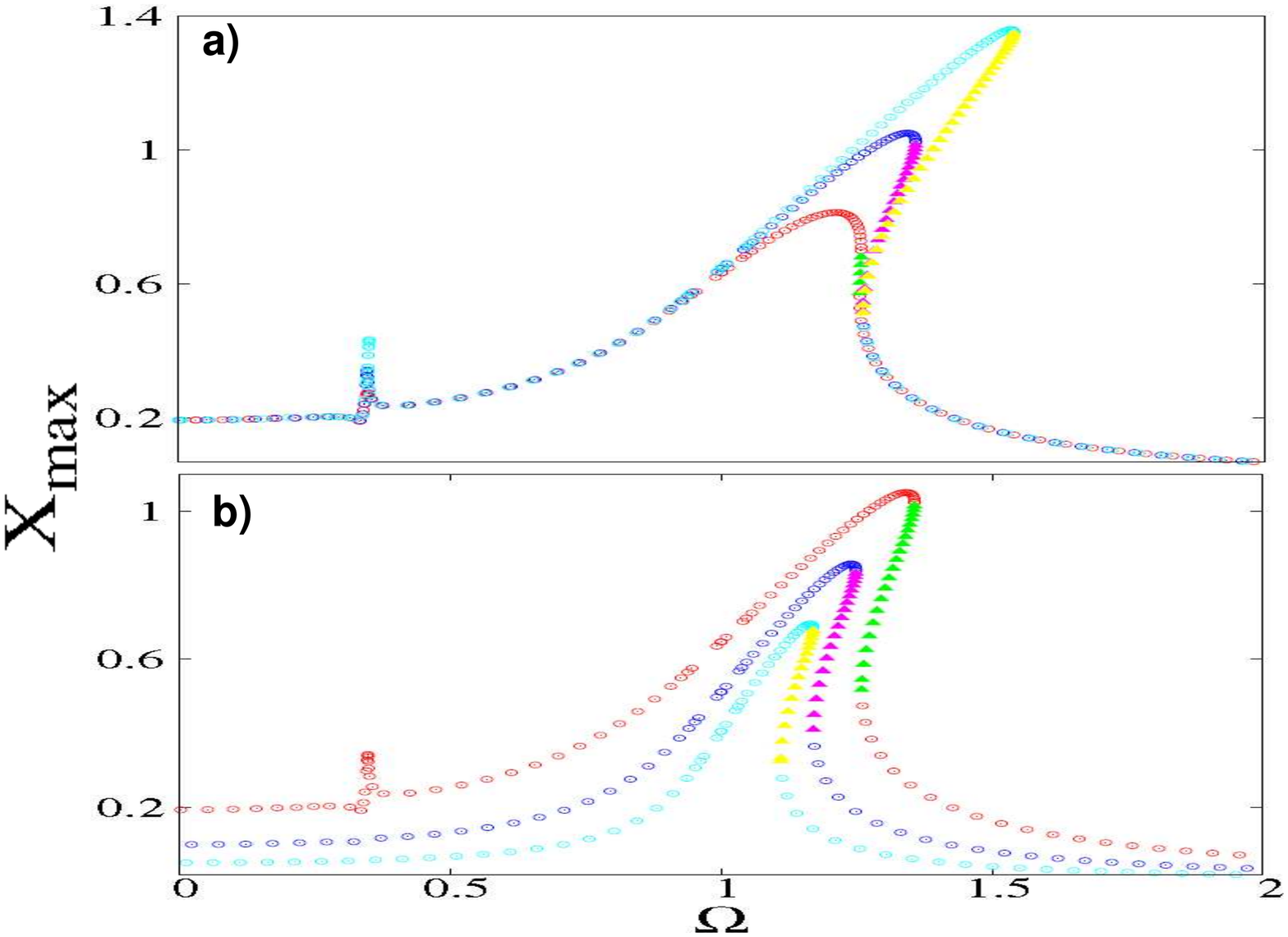}
\caption{(Color online) Maximum oscillation amplitude, $|x|$, as a function of $\Omega$ in the case of zero noise: (a) for a fixed value of the driving force $F_D=0.2$ and different values of $\eta$, from bottom to top $\eta= 1.2$, $\eta=0.5$, $\eta=0.2$; (b) for a fixed value of the nonlinear damping coefficient $\eta=0.5$ and different values of $F_D$ from bottom to top $F_D=0.05$, $F_D=0.1$, $F_D=0.2$. Empty circles correspond to stable orbits and solid triangles to unstable orbits.}
\label{etavsomega}
\end{figure}

Duffing resonators present a characteristic resonant line shape. They have a resonant peak for the maximum oscillation amplitude $|x|$ at the driving angular frequency $\Omega=\Omega_{res}$ and a force-dependent resonance width $\Delta \Omega$. The latter is the difference between the frequencies of the minima flanking the resonant peak, and it is constant for linear resonators. When the changes to the resonance due to dissipation or noise are smaller than the resonance width, the standard definition of quality factor is $Q= \omega_0\tilde{E}/\langle d\tilde{E}/d\tilde{t}\rangle = 1.09\,\Omega/\Delta\Omega$ ($\tilde{E}$ the mechanical energy at a given time and $\langle\ldots\rangle$ denotes time-averaging over a time scale long compared with the oscillation period but sufficiently short that the decay of the amplitude is negligible). See the Supplementary material in Ref.~\onlinecite{Bach}. For linear resonators the resonant frequency $\Omega_{res}$ and the quality factor, $
 Q$, are independent of the driving force \cite{Bach}, whereas for nonlinear resonators the energy dissipation and hence the quality factor depend on the oscillation amplitude. Numerical simulations show that the resonant frequency $\Omega_{res}$ increases with increasing $\eta$ (Figure \ref{etavsomega}a) and $F_D$ (Figure \ref{etavsomega}b). The width of the characteristic peak also depends on $\eta$ and $F_D$.

The dependence of $Q$ on the driving force and on the nonlinear dissipation term is not the only characteristic of Duffing resonators. One important feature of these systems is the presence of hysteresis in the oscillation amplitude as a function of the driving frequency or the driving force.\cite{BoA} Figure \ref{etavsomega} shows that the system has more than one possible oscillation amplitude for a range of values of $\Omega$. In the bistable region there are two stable fixed points corresponding to periodic orbits in the case of periodically driven wires, one of larger oscillation amplitude (higher in energy) and one of smaller oscillation amplitude (lower energy). There is also one unstable periodic orbit located between the stable ones. The dark line (blue line in the color version) on Figure \ref{etavsomega} corresponds to unstable periodic orbits while the light (green) line corresponds to the stable ones.

\begin{figure}
\includegraphics[scale=0.5]{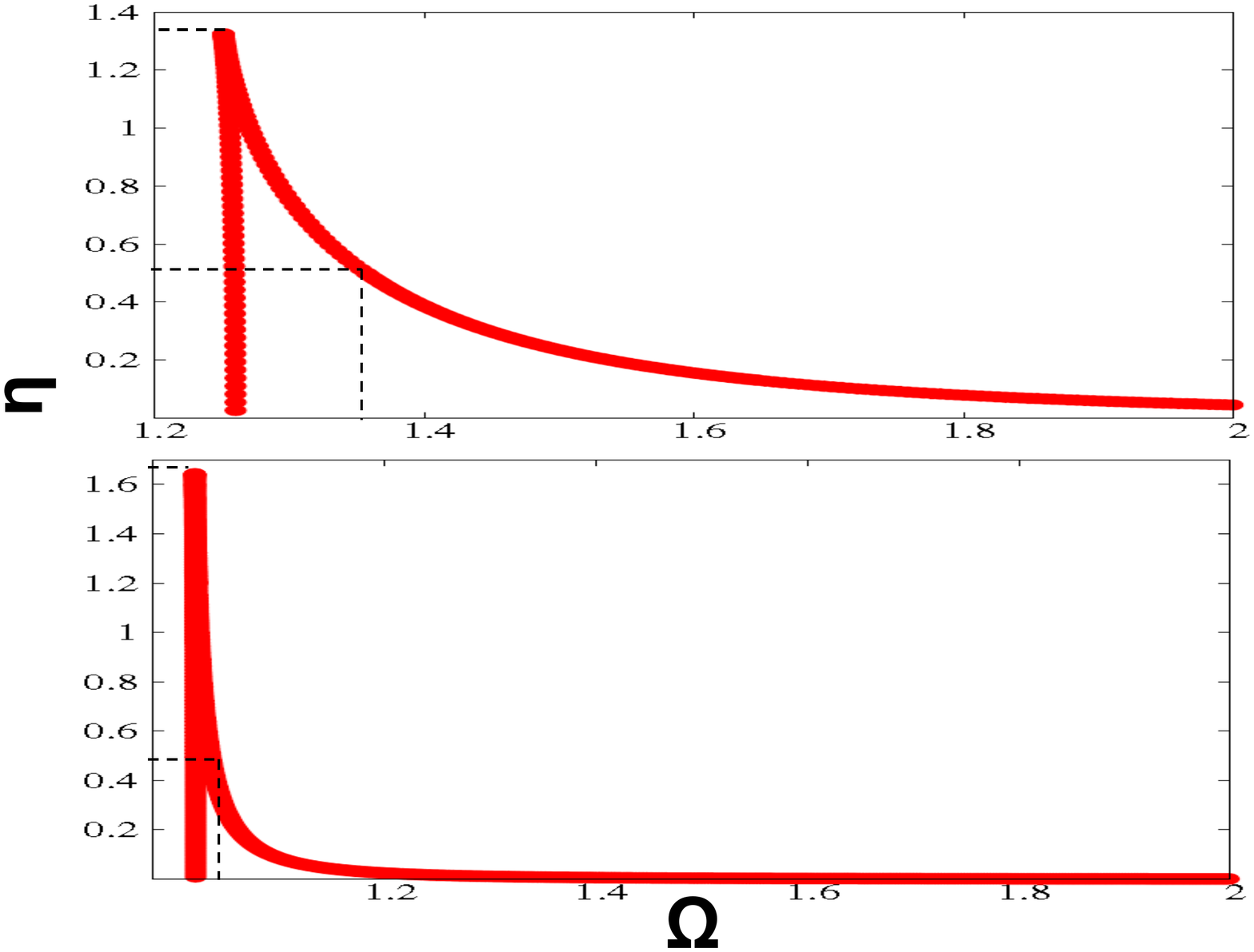}
\caption{(Color online) Bistability zone as function of nonlinear dissipation amplitude $\eta$ and $\Omega=2\pi f$ for fixed values of the driving amplitude: (a) $F_D=0.2$ and (b)$F_D=3.8\times10^{-3}$. The projection of the red curve on the $\Omega$ axis represents the minimum and maximum values of $\Omega$ at which the system is bistable. The size of the hysteresis window, $\Delta H=\Omega_{max}-\Omega_{min}$, depends on both the driving force and the nonlinear dissipation parameter [compare panels (a) and (b)]. For example, for $\eta=0.5$, the minimum (maximum) frequency at which hysteresis can be observed is $\Omega_{min}=1.262$ (resp.\ $\Omega_{max}=1.357$) in panel (a) and $\Omega_{min} =1.02$ (resp.\ $\Omega_{max}=1.034$) in panel (b). As $\eta$ increases, $\Delta H$ gets reduced until the inflection point, $\Omega_{max}=\Omega_{min}$, at which hysteresis completely disappears.}
\label{etavsomega2}
\end{figure}

The two key elements in the hysteretic behavior, discussed so far in both theory and experiments, are the Duffing term, responsible for the system bistability, and the nonlinear dissipation term $\tilde{\eta} \tilde{x}^2\dot{\tilde{x}}$, which controls the size of the hysteresis region. In the limit of weak linear damping, perturbation theory based on the small parameter $\epsilon=Q^{-1}\ll 1$ shows that it is necessary to have $\eta<\sqrt{3}$ for the solutions of equation (\ref{dimless}) to exhibit bistability.\cite{lifshitz} However this condition only provides an upper bound. Numerical simulations show that the maximum $\eta$ for which bistability is present, $\eta_M$, depends on the driving force amplitude $F_D$. Figures \ref{etavsomega2}a and \ref{etavsomega2}b depict $\eta_M$ as a function of $F_D$ and the size of the hysteresis window (the range of frequencies at which there is hysteresis), $\Delta H$, as a function of $\eta$. Note that $\eta_M(0.2)=1.325$, indicated i
 n Fig.~\ref{etavsomega2}a, is appreciably smaller than the upper bound $\sqrt{3}$, whereas for a much smaller $F_D=3.8\times10^{-3}$, $\eta_M=1.642$ as indicated in Fig.~\ref{etavsomega2}b is closer to $\sqrt{3}$. Increasing $\eta$ reduces the frequency range at which the system is bistable. For a given driving force $F_D$, the minimum frequency $\Omega_{min}$ at which there is bistability does not vary too much, whereas the maximum frequency $\Omega_{max}$ at which hysteresis can be observed  moves toward smaller values. In Figure \ref{etavsomega}a, $\Omega_{min}=1.257$ and $\Omega_{max}$ is seemingly unbounded for $\eta=0$. For $\eta=0.5$, $\Omega_{min}$ has moved slightly to $\Omega_{min}=1.251$ while the maximum frequency has been considerably reduced to $\Omega_{max}=1.357$. This reduction continues until $\eta$ reaches a critical point where $\Omega_{min}=\Omega_{max}$ and bistability completely disappears. This happens at $\eta=1.325$ in the case of $F_D=0.2$ (Figure
 \ref{etavsomega2}a), and at $\eta=1.642$ for $F_D=3.8\times10^{-3}$ (Figure \ref{etavsomega2}b). Eichler \emph{et al} estimate $\tilde{\eta}=7.9\times10^{5}$ kg/(m$^{2}$s) $>\alpha\sqrt{3}/\omega_0$ (corresponding to $\eta>\sqrt{3}$) for carbon nanotubes and $\tilde{\eta}=1.5\times10^{5}$ kg/(m$^{2}$s) $<\alpha\sqrt{3}/\omega_0$ ($\eta<\sqrt{3}$) for graphene resonators. \cite{Bach}

So far the only parameter controlling the size of the hysteresis region or the dependence of $Q$ on the driving force has been the nonlinear dissipation coefficient $\eta$. This coefficient is an intrinsic property of each material and is fixed for every set up. However there is a parameter that may give rise to similar effects and that can be tuned or controlled externally. In the next section we explore the effect that the addition of noise has on the dynamics of the system and show that it has similar properties as those found for $\eta$.
\begin{figure}
\begin{center}
\includegraphics[scale=0.6]{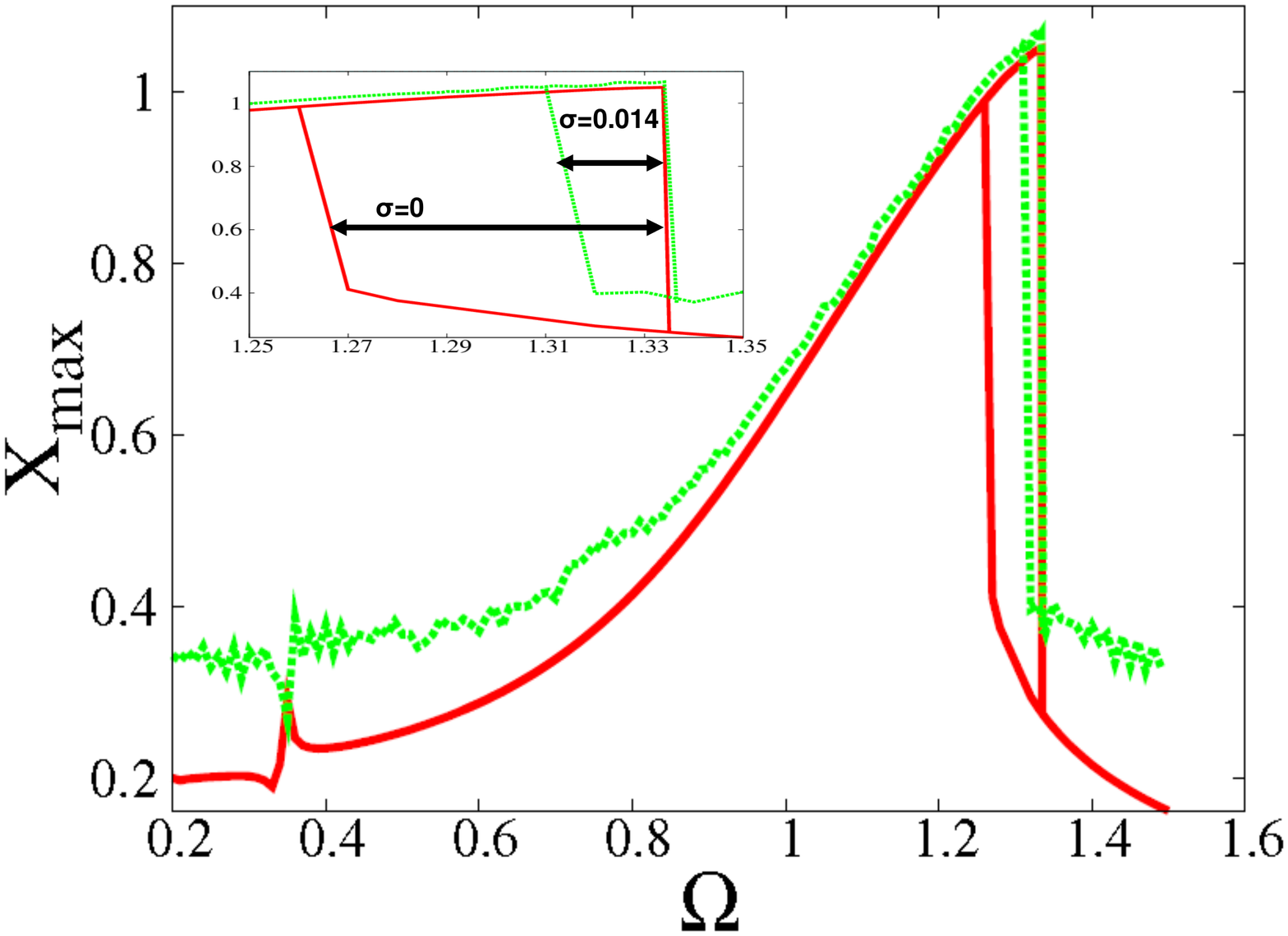}
\end{center}
\caption{(Color online) Curve of the maximum oscillation amplitude as a function of $\Omega$ at zero noise, (solid line) and at $\sigma=0.014$ (doted line) for a given  $\eta=0.5$, $F_D=0.2$, $Q^{-1}=5\times10^{-4}$. The area under the big solid square corresponds to the bistability zone at zero noise, and the area under the smaller doted square represents the bistability zone for a finite noise amplitude $\sigma$}
\label{hyst:plot}
\end{figure}
\section{ External noise: Results and Discussion}\label{sec:3}
\begin{figure}
\includegraphics[scale=0.55]{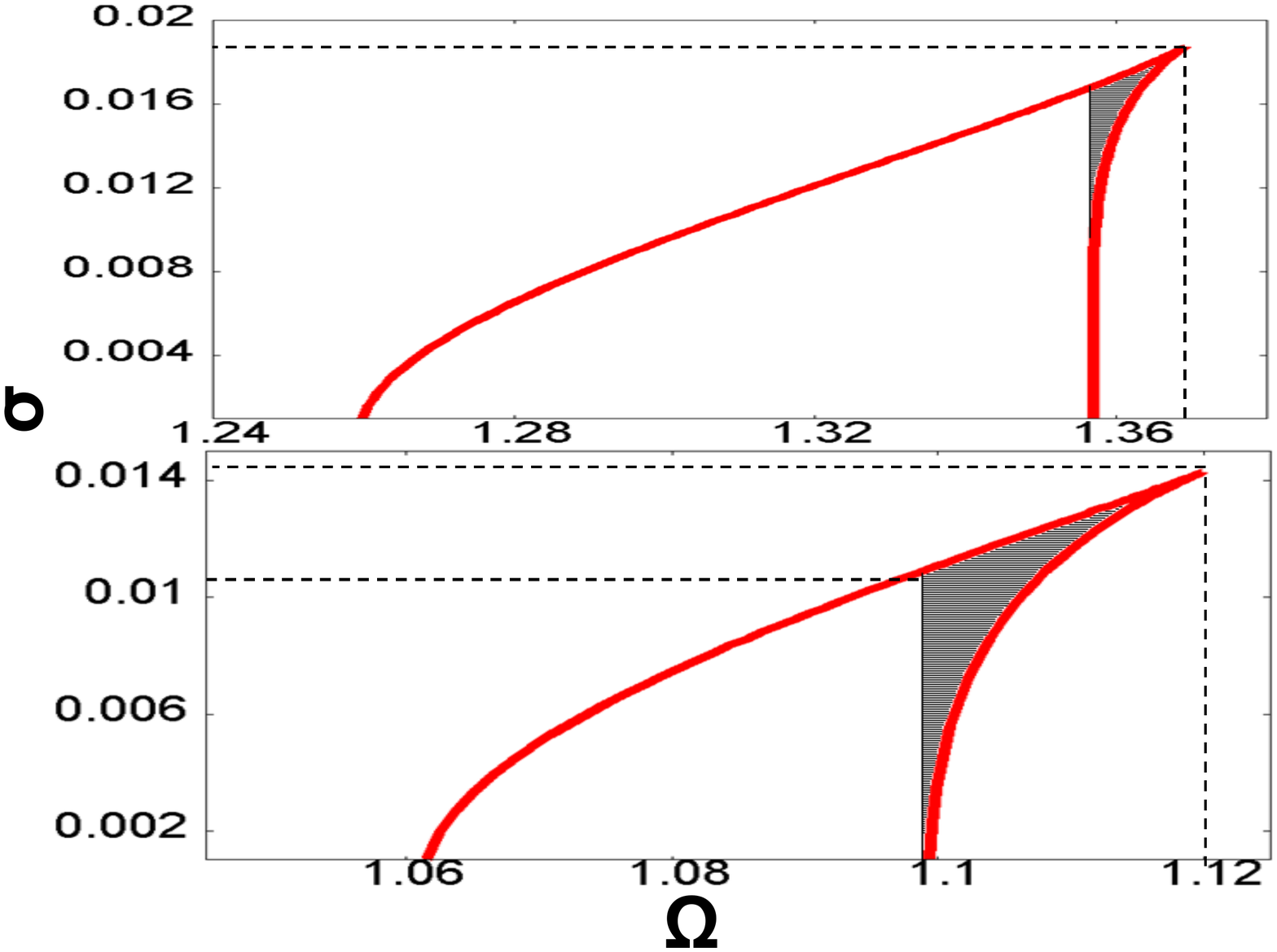}
\caption{(Color online) Bistability zone in the plane $(\sigma,\Omega)$ for fixed nonlinear damping $\eta=0.5$ and (a) $F_D=0.2$, (b) $F_D=.02$. The projection of the curve on the $\Omega$ axis gives the minimum and maximum frequencies at which there is hysteresis. The shaded zone corresponds to values of $\Omega$ at which there is no hysteresis for $\sigma=0$. This region gets smaller as the noise is increased. Comparing (a) and (b), we can observe that this region also gets smaller for smaller values of $F_D$}
\label{Diff_sigma}
\end{figure}

\begin{figure}
\includegraphics[scale=0.55]{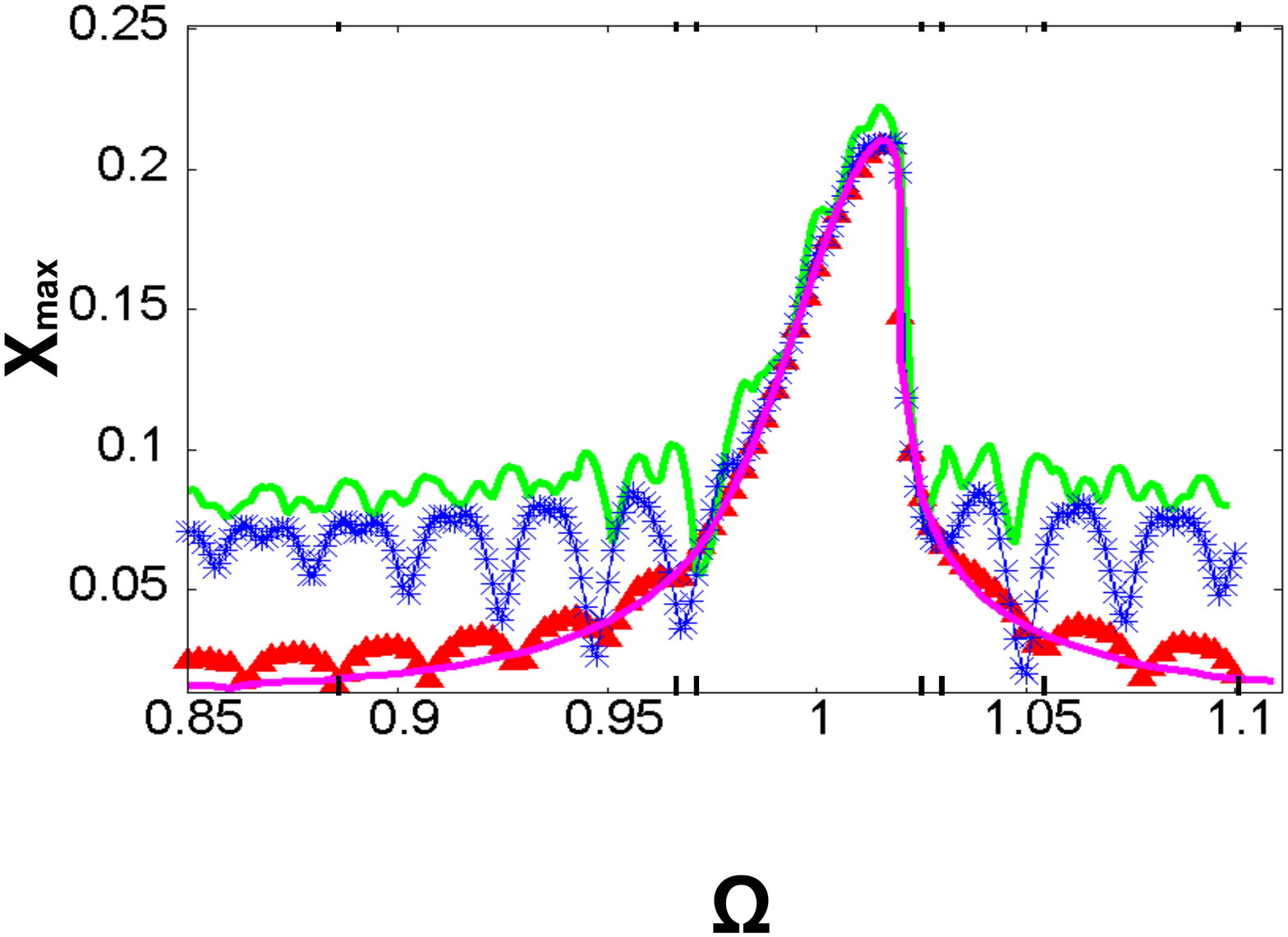}
\caption{(Color online) Maximum oscillation amplitude $X_{max}$ as a function of $\Omega$, for $\eta=1.6$ and $F_D=0.0038$. bottom solid line correspond to zero noise and top solid line $\sigma=F_D$, in this regime hysteresis has completely disappeared. The last minimum before the curve increases monotonically to its global maximum are located at $\Omega=1.1$ for $\sigma=0$(bottom solid line), at $\Omega=1.055
$ for $\sigma=0.05F_D$ (solid triangles), and for $\sigma=0.5F_D$ (stars) and at $\Omega=0.97$ for $\sigma=F_D$ (top solid line). The first minimum before the curve decreases are located at $\Omega=0.885$ for $\sigma=0$(bottom solid line), at $\Omega=1.03$ for $\sigma=0.05F_D$ (solid triangles), and for $\sigma=0.5F_D$ (stars) and at $\Omega=1.026$ for $\sigma=F_D$ (top solid line) }
\label{width}
\end{figure}

According to the fluctuation-dissipation theorem, noise accompanies the process of energy dissipation. Thus we can expect similar behaviors as those related to $\eta$ when noise is taken into account. For such purpose we add an external white noise to equation (\ref{dimless})
\begin{eqnarray}\label{noise:eq}
&&\ddot {x}=-x-Q^{-1}\dot {x}-x^3-\eta x^2\dot {x} + F_D\cos(\Omega t)+\sigma \xi(t),\\
&&\langle \xi(t)\rangle =0,\quad \langle \xi(t)\xi(t')\rangle = \delta(t-t'),  \nonumber
\end{eqnarray}
which is presumably larger than the intrinsic noise. Numerical simulation of this stochastic equation shows that the noise shrinks the hysteresis window (Figures \ref{hyst:plot}, \ref{Diff_sigma}a,b) and reduces the resonance width (Figure \ref{width}).

For fixed values of $F_D$ and $\eta$, the hysteresis window is reduced by increasing the noise strength, $\sigma$. Figure \ref{hyst:plot} shows that, at $\eta=0.5$, the hysteresis window has been reduced from $\Delta H=0.098$ at $\sigma=0$ to $\Delta H=0.04$ at $\sigma=0.014$. In contrast with the effect of increasing $\eta$, increasing $\sigma$ increases fast the minimum frequency $\Omega_{min}$ at which bistability starts, whereas the maximum frequency $\Omega_{max}$ is smaller than that of the deterministic case but it does not change than much with $\sigma$; see Figure \ref{Diff_sigma}. Let us now compare the effects of $\eta$ and $\sigma$ on the hysteresis window. In Figure \ref{etavsomega2}, we decrease the window size by increasing $\eta$ from zero to $\eta=0.7671$ at zero noise, whereas in Figure \ref{Diff_sigma} we do so by increasing $\sigma$ from zero to $\sigma=0.014$ at $\eta=0.5$. In both cases $\Delta H=0.04$ is the same, however $\Omega_{max}$ and $\Omega_{min
 }$ are different: (i) $\Omega_{max}=1.299$, $\Omega_{min}=1.259$ for $\eta=0.7671$ and $\sigma=0$, and (ii) $\Omega_{max}=1.318$, and $\Omega_{min}=1.278$ for $\sigma=0.014$ and $\eta=0.5$. The hysteresis window has shifted towards higher values of $\Omega$. The critical value at which hysteresis disappears, $\Omega_{max}=\Omega_{min}$, occurs at $\Omega= 1.251$ for $\sigma=0$ and $\eta=1.325$ (Figure \ref{etavsomega2}). However, for $\eta=0.5$ and $\sigma=0.014$, it occurs at $\Omega=1.369$ far above the original zone of hysteresis at zero noise (shaded area on Figure \ref{Diff_sigma}).  Summarizing, increasing $\sigma$ not only reduces the hysteresis window (as it has already been found experimentally),\cite{Aldridge} it also shifts the window towards higher frequencies. This is so to the point that, just before the critical frequency $\Omega_{max}=\Omega_{min}$, hysteresis is found on a frequency region in which there is no bistability at zero noise (shaded area on Figure
 s \ref{Diff_sigma}a and \ref{Diff_sigma}b).

The fact that the hysteresis window moves towards higher frequency values when $\sigma$ increases is more than just a curiosity. It is telling us a lot about what is going on in the system: it means that noise is feeding energy into the system. This can be understood with the following argument. In the hysteretic region there are three periodic orbits, one unstable orbit separates stable orbits with larger and smaller amplitude (therefore with higher and lower energy), as seen in Figures \ref{etavsomega} and \ref{hyst:plot}. When the hysteresis zone is reduced, only one of the stable orbits survives. The other stable orbit coalesces with the unstable one and both then disappear. As  $\eta$ increases the surviving orbit has the smaller amplitude: as $\eta$ increases, the system is dissipating more energy and its dynamics selects lower energy orbits. In contrast to this, the larger amplitude orbit survives when $\sigma$ increases because the noise is feeding energy to the syste
 m making more likely for the system to oscillate around the higher energy periodic orbit. It also turns out that the quality factor increases with noise which enforces this argument.

Even more important than the reduction of $\Delta H$ is that increasing the noise strength makes the resonance width slightly thinner. In the absence of noise the two minima flanking the peak are hardly perceptible. The two deep minima observed in the experiments of Eichler et al in Ref. \onlinecite{Bach} make us suspect that noise was present in those experiments. This effect that resonance width shrinks with noise  is always present, even in systems not exhibiting hysteresis. When $\sigma$ increases, the two minima flanking the peak become closer and deeper. On Figure \ref{width}, we see how the two minima are almost imperceptible and are located at $\Omega=0.88$ and $\Omega=1.05375$ at zero noise (green crosses). For $\sigma=F_D$, the two minima are more pronounced and are located at $\Omega=0.9765$ and $\Omega=1.03125$. The resonance width has been reduced from $\Delta\Omega=0.173$ at $\sigma=0$ to $\Delta\Omega=0.05$ at $\sigma=F_D$ (blue stars). This shrinking can be se
 en in Figure $\ref{width}$ for more values of $\sigma$. This indicates that the presence of noise is affecting the value of the quality factor $Q$. Since the reduction/enlargement due to noise and dissipation are smaller than the resonance width, the quality factor is\cite{Bach}
\begin{equation}
Q=\Omega_0 \frac{\langle E\rangle}{\langle \dot{E}\rangle}\approx 1.09\, \frac{\Omega}{\Delta\Omega}, \label{q-factor}
\end{equation}
where $E$ is the mechanical energy at a given time $t$. According to the Supplementary material in Ref.~\onlinecite{Bach}, the approximation $Q\approx 1.09 \Omega/\Delta\Omega$ holds in the limit of very large quality factor. The noise reduces $\Delta \Omega$ and therefore it enhances the quality factor. This contrasts with the behavior of linear systems. In them, external noise increases energy dissipation and makes these systems less efficient. However for our nonlinear system, the random energy provided by noise gets absorbed by the system and converted in useful mechanical energy, thereby yielding a higher quality factor. In this sense, the nonlinearly damped driven Duffing oscillator is a {\em noise rectifier} that improves its performance by using energy provided by the external noise.

The reduction of the frequency width is also telling us that the value of $\eta$ calculated from the experiments of Eichler \emph{et al}\cite{Bach} (in which noise is certainly present) may be smaller than the real one, i.e., $\eta|_{\sigma=0}>\eta|_{\sigma>0}$. The value of $\eta$ in Ref.~\onlinecite{Bach} is calculated from the expression
\begin{equation}
\eta= \left(\frac{m\,\Delta f}{0.032}\right)^3\left(\frac{F_D}{f_0}\right)^2,\label{eta}
\end{equation}
where the dependence of $\Delta\Omega$ on $\sigma$ has not been taken into account. We can include this dependence on equation (\ref{eta}) and let $\eta(\sigma)$ be given by (\ref{eta}) for a fixed value of $\sigma\geq 0$. We can rewrite $\Delta \Omega= \Delta \Omega\mid_{\sigma=0}/f(\sigma)$, with $f(\sigma) = (\Delta\Omega\mid_{\sigma=0}/\Delta \Omega\mid_{\sigma>0})$. The factor $f(\sigma)$ indicates how the resonance width changes with noise: $f(0)=1$ and $f(\sigma)>1$ for $\sigma>0$. Using (\ref{eta}), we see that
\begin{equation}
\eta(\sigma)=\frac{\eta(0)}{f(\sigma)^3}\leq \eta(0).\label{compare_eta}
\end{equation}
This means that the estimated value $\eta(\sigma)$ is smaller than the real nonlinear dissipation coefficient of the deterministic equation, $\eta=\eta(0)$. The real value of $\eta$ is screened by the effect of noise.

\begin{figure}
\includegraphics[scale=0.55]{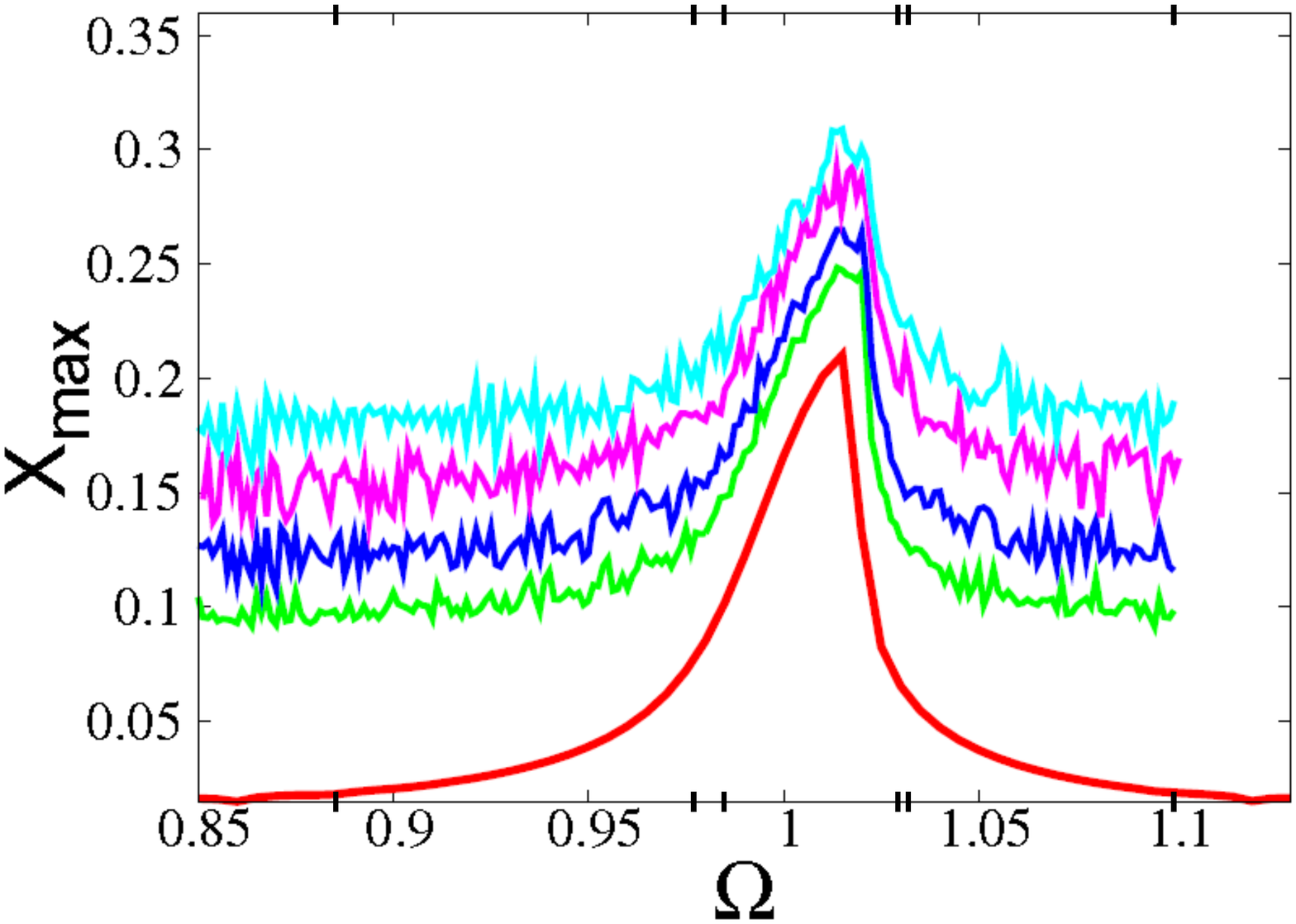}
\caption{(Color online) Maximum oscillation amplitude $X_{max}$ as a function of $\Omega$, for $\eta=1.6$ and $F_D=0.0038$ when the noise strength is given by the fluctuation-dissipation theorem, $\sigma^2= 2\delta(\eta x^2 +1/Q)$, with $\delta =\alpha k_B T/(k^2)$. From bottom to top $T=0$, $T=50$K, $T=100$K, $T=200$K, $T=300$K. It can be observed that the resonance width is reduced as $T$ (hence also $\sigma$) increases, although in this case it is more difficult to identify the two minima flanking the peak.}
\label{Temp}
\end{figure}

\section{Internal noise}\label{sec:4}
So far we have discussed a general case of external white noise, using a constant noise amplitude $\sigma$, without specifying the nature of it. However, in our system there are different sources of dissipation that contribute to $\gamma$ and $\tilde{\eta}$ in (\ref{eq1}). Let us assume that the dominant noise is thermo-mechanical.\cite{sazonova} According to the fluctuation-dissipation theorem, there should be a stochastic white noise forcing term in (\ref{noise:eq}) with an appropriate strength, $\sigma^2= 2 \delta (\eta x^2+1/Q)$, where $\delta=\alpha k_BT/k^2$. We have performed numerical simulations of (\ref{noise:eq}) with this internal noise for a given value of $\eta$. The results obtained are qualitatively similar to those discussed above. Figure \ref{Temp} shows that increasing the temperature reduces the resonance width (similar to the effect of increasing the fixed strength $\sigma$ of an external noise), thereby increasing the quality factor. Similarly, the hyste
 resis frequency range diminishes with all the consequences already mentioned.

The measurements of the quality factor made by Eichler \emph{et al} are not affected by these results, since they calculate $Q=1.09\,\Omega_0/\Delta \Omega$ and take $\Delta\Omega$ from the experimental measurements which are affected by the existing noise. However the inferred value of $\tilde{\eta}$ may be affected by including noise. The measurements in Ref.~\onlinecite{Bach} were interpreted assuming that only the nonlinear dissipation $\tilde\eta$ was responsible for the shrinking or expansion of the resonant peak. Since noise also has an effect, the value of $\tilde\eta$ proposed in Ref.~\onlinecite{Bach} could be smaller than the real one.

\section{Conclusions}\label{sec:5}
We have found that the presence of noise modifies in a qualitative way the dynamics of the nonlinear Duffing resonators, and it gives rise to some effects similar to those attributed to the nonlinear dissipation coefficient. Noise not only reduces the window of hysteresis but also shifts it to higher values of the driving frequency. Since the value of $\eta$ is inherent to the material and can not be tuned, there is not much room for modifying the quality factor rather than increasing $F_D$ which cannot always be done experimentally. However controlling the sources of noise allows to increase or decrease the bistability regions and even to shift them to higher frequencies within easy experimental reach.

We have also found that the resonant width can be controlled by noise and hence the quality factor can be increased by increasing the sources of noise in the system. This means not only that higher quality factors can be reached by increasing the sources of noise, but also, and even more important, it means that nanoresonators act as noise rectifiers. This is the main result of this work, to show that nonlinear nanoresonators are able to convert random energy from the environment into useful mechanical energy.

Even though there are different sources of noise present in these systems,\cite{Cleland,sazonova} in this work we have discussed mostly the case of white noise with constant strength $\sigma$, leaving its nature unspecified. However, we also performed numerical simulations assuming that the dominant noise is of thermo-mechanical origin and satisfies the fluctuation-dissipation theorem (the noise strength $\sigma = \sqrt{2\delta(\eta x^2 +1/Q)}$, with $\delta =\alpha k_B T/k^2$, depends on the temperature and on the oscillation amplitude). The results are qualitatively similar to those for constant strength white noise.

\acknowledgments
This work has been supported by the Spanish Ministerio de Econom\'\i a y Competitividad grants FIS2011-28838-C02-01, FIS2010-22438-E (Spanish National Network Physics of Out-of-Equilibrium Systems), MAT 2011-24331 and ITN, Grant No. 234970 (EU) .

\end{document}